\documentclass[12pt]{iopart}
\usepackage{amssymb}
\usepackage{graphicx}
\usepackage{dcolumn}
\usepackage{enumerate}
\usepackage{bm,bbm}

\expandafter\let\csname equation*\endcsname\relax
\expandafter\let\csname endequation*\endcsname\relax

\usepackage{amsmath}

\begin{document}

\title[Transport enhancement from incoherent coupling between one-dimensional quantum conductors]{Transport enhancement from incoherent coupling between one-dimensional quantum conductors}

\author{J J Mendoza-Arenas$^1$, M T Mitchison$^{2,1}$, S R Clark$^{3,1}$, J Prior$^4$, D Jaksch$^{1,3}$ and M B Plenio$^{5,2}$}
\address{$^1$Clarendon Laboratory, University of Oxford, Parks Road, Oxford OX1 3PU, United Kingdom}
\address{$^2$Quantum Optics and Laser Science Group, Blackett Laboratory, Imperial College London, London SW7 2BW, United Kingdom}
\address{$^3$Centre for Quantum Technologies, National University of Singapore, 3 Science Drive 2, Singapore 117543}
\address{$^4$Departamento de F\'{i}sica Aplicada, Universidad Polit\'{e}cnica de Cartagena, Cartagena 30202, Spain}
\address{$^5$Institut f\"{u}r Theoretische Physik, Albert-Einstein Allee 11, Universit\"{a}t Ulm, 89069 Ulm, Germany}

\ead{j.mendoza-arenas1@physics.ox.ac.uk}

\begin{abstract}
We study the non-equilibrium transport properties of a highly anisotropic two-dimensional lattice of spin$-\frac{1}{2}$ particles governed by a Heisenberg $XXZ$ Hamiltonian. The anisotropy of the lattice allows us to approximate the system at finite temperature as an array of incoherently coupled one-dimensional chains. We show that in the regime of strong intrachain interactions, the weak interchain coupling considerably boosts spin transport in the driven system. Interestingly, we show that this enhancement increases with the length of the chains, which is related to superdiffusive spin transport. We describe the mechanism behind this effect, compare it to a similar phenomenon in single chains induced by dephasing, and explain why the former is much stronger.
\end{abstract}

\section{Introduction} \label{intro}

Since the experimental discovery that quantum coherent dynamics is present in excitation energy transport in biological light harvesting complexes \cite{engel2007nat} and theoretical work demonstrating that environmental fluctuations can be used to optimise transport efficiency \cite{plenio2008njp,mohseni2008jcp}, a great deal of interest has focused on the beneficial consequences that the unavoidable coupling of a quantum system to its environment can have \cite{noise_enhanced_capacity,deph_assisted_plenio2, plenio1999pra, Krauter_prl2011,barreiro2010nat,simulation_optical_cavities,kendon2003pra,javier2013nat}. In particular, it has been found that the optimal regime for excitation transport through various systems consists of a balance between coherent and incoherent phenomena, or, more specifically, the interplay between coherent electronic dynamics and the vibrational environment \cite{plenio2008njp,mohseni2008jcp,deph_assisted_plenio,huelga2013cp}. Most of this effort to characterise and understand environment-assisted transport has been restricted to single-particle effects \cite{plenio2008njp,mohseni2008jcp,deph_assisted_plenio,nazir2009prl,kassal2012njp}, given that light-harvesting complexes under physiological conditions usually contain very few excitations at the same time due to the low intensity of ambient sunlight \cite{ishizaki2010pccp,scholes2011nat}. Nevertheless, since the interplay between coherent and incoherent phenomena is relevant beyond a biological context, it is important to consider its impact on the transport properties of many-body systems.

\begin{figure}
\begin{center}
\includegraphics[scale=0.7]{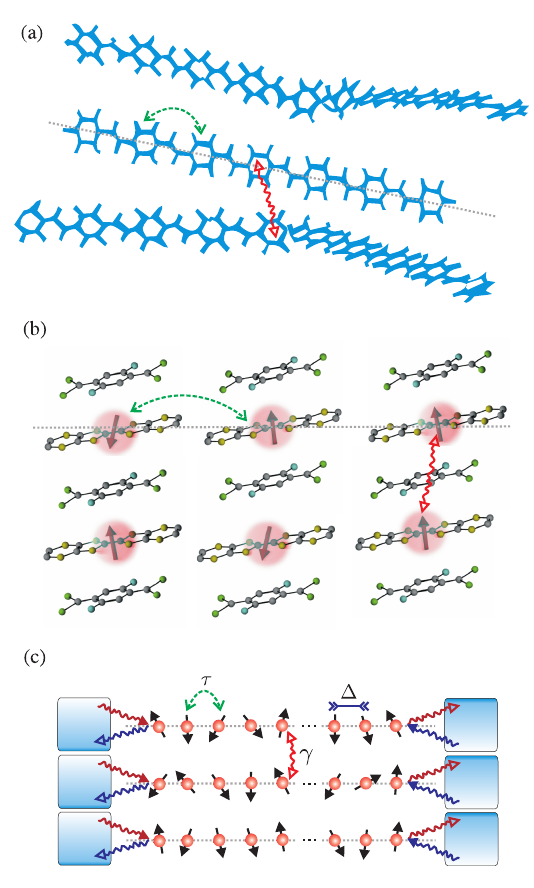}
\caption{\label{fig1} Illustration of systems of incoherently coupled chains. The red solid and green dashed lines represent interchain and intrachain hopping, respectively. (a) Excitation transport in a conjugated polymer system. Interchain hopping of excitons, observed to be incoherent \cite{collini2009sci,collini2009jpca}, occurs where polymeric chains are close to each other. (b) Excitation hopping in organic salts. Coherent hopping occurs along one direction of the system, while incoherent hopping takes place along another direction, mediated by a scaffolding of different molecules. (c) System of three incoherently coupled spin chains, with intrachain hopping $\tau$, interaction strength $\Delta$ and interchain coupling $\gamma$. The blue arrows represent the right-to-left driving of the system, while the red arrows correspond to the left-to-right driving.}
\end{center}
\end{figure}

It has been known for many years that coherent and incoherent particle transport processes take place in various condensed matter systems. These include cuprates \cite{turlakov2001prb,levin2004prb,vignolle2012prb} and several organic conductors such as conjugated polymers \cite{collini2009sci,collini2009jpca}, layered organic metals \cite{kartsovnik2006prl,kartsovnik2009prb} as well as Bechgaard and Fabre salts \cite{jerome1982advphys,dressel2005prb,kohler2011prb}. As illustrated in figures~\ref{fig1}(a),(b), these systems share a common feature, which is a highly anisotropic structure consisting of lattice sites strongly coupled in one or two directions and weakly coupled in other directions. At not-too-low temperatures, particle transport along the strongly coupled directions is predominantly coherent, while the transport along directions with weak coupling occurs via incoherent hopping processes. This simple picture is incomplete, however. There exist several competing effects that may contribute to the transport behaviour, depending on the material in question. These include static disorder, interparticle interactions, local dissipation, and spatially correlated noise. It is thus expected that a rich variety of phenomena emerges from the interplay between these different processes.

In order to disentangle the contributions from these various effects, a natural strategy is to analyse the competition between just a few of them first, which can be non-trivial. For instance, it is well known that static disorder results in Anderson localisation, which is broken by weak dissipation or decoherence effects~\cite{plenio2008njp,znidaric2010dephasing}. On the other hand, the interplay between interactions and disorder is not fully understood, and is still the object of intense research~\cite{basko2006ann,albrecht2012prb,lucioni2011prl}. In the present work, we focus exclusively on the physics resulting from the combination of coherent interactions and incoherent processes, neglecting disorder and other complications. This sets the stage for a more complete future analysis that includes all of these elements. 


Several interesting effects resulting from the competition between coherent and incoherent processes in the presence of strong interactions have recently been found \cite{pichler2010pra,kollath2012prl,cai2013prl}. In particular, previous research by some of us has uncovered a novel mechanism of dephasing-enhanced transport in linear homogeneous strongly interacting systems \cite{we,we2}. Nevertheless, the physics resulting from this interplay still represents relatively uncharted territory. 

Thus, motivated by the existence of coherent and incoherent hopping processes in several anisotropic condensed matter systems, we propose a concrete minimal model that contains these features, as shown in figure \ref{fig1}(c). Specifically, we consider excitation transport through arrays of incoherently coupled one-dimensional quantum spin chains, including the possibility of strong interactions between excitations on the same chain. We study DC transport properties by imposing a net current flowing in one direction through the system. We find that the effective environment furnished by nearby chains significantly enhances intrachain transport for sufficiently strong interactions between excitations. In addition, such a transport enhancement increases with the size of the chains, indicating the relevance of the mechanism for bulk materials. Furthermore, the incoherent hopping of spin excitations between chains results in a much more pronounced effect than that produced by pure dephasing due to, for example, lattice vibrations \cite{we,we2}. We emphasise that the simple model we consider does not account for several effects, e.g.\ dissipation and disorder, expected to be relevant for real systems such as organic conductors and cuprates. Nevertheless the results we present in this work are general and do not depend on the particular form of the interaction. Therefore we believe that they constitute a meaningful contribution towards the understanding of the rich phenomenology arising from the interplay between coherent and incoherent effects.

The paper is organised as follows. In Section \ref{model} we describe the model to be studied and the approximations considered. In Section \ref{weak} we study weakly interacting systems, where the incoherent coupling only degrades the transport. In Section \ref{subassisted} we analyse the case of strong interactions, where current enhancement due to incoherent coupling is observed. The origin of this effect is explained, and compared to that resulting from dephasing processes \cite{we,we2}. Finally, our conclusions are discussed in Section \ref{conclu}.

\section{Model of non-equilibrium incoherently coupled spin chains} \label{model}

In this work we consider a $N\times \Lambda$ rectangular two-dimensional (2D) lattice, consisting of $\Lambda$ chains with $N$ sites each (see figure~\ref{fig1}(c)). The model for coherent intrachain transport should describe conserved excitations that can hop between lattice sites and interact with each other. We therefore choose the simple spin-$\frac{1}{2}$ $XXZ$ Hamiltonian to govern the dynamics of each chain \cite{gobert2005real,langer2009real}
\begin{equation} \label{hami}
H^{(\lambda)} = \sum_{i=1}^{N-1}\tau (\sigma_i^{x(\lambda)}\sigma_{i+1}^{x(\lambda)}+\sigma_i^{y(\lambda)}\sigma_{i+1}^{y(\lambda)}+\Delta \sigma_i^{z(\lambda)}\sigma_{i+1}^{z(\lambda)}),
\end{equation}
where the super-index $(\lambda)$ refers to any operator of chain $\lambda$, $\sigma^{k(\lambda)}_i$ ($k=x,y,z$) are Pauli matrices at lattice site $i$ of chain $\lambda$, $\tau$ is the exchange coupling between nearest neighbours (in the following we take units of energy and time such that $\tau=1$ and $\hbar  = 1$), and $\Delta$ is the anisotropy (we consider $\Delta>0$ only), where both parameters are assumed to be the same for every chain. The presence of an excitation at a certain site corresponds to a spin pointing up, while the absence of an excitation corresponds to a spin pointing down. The hopping is encapsulated by the first two terms of equation~\eqref{hami}, while the final term corresponds to an energy penalty for nearest-neighbour lattice sites in the same spin state, creating an interaction between spin excitations. In this sense, we will refer to a strongly interacting model if $\Delta>1$, which is an energy-gapped regime.

We assume that hopping also occurs between nearest-neighbour sites of neighbouring chains, with a hopping rate $\eta$. In addition, we suppose that the interchain coupling is much weaker than the coupling between sites on the same chain, so that $\eta\ll\tau$. Let $t_{\phi}$ denote the time taken for a spin excitation to lose its phase coherence, either from collisions with other spin excitations on the same chain or due to dephasing induced by an external bath, e.g.\ phonons. If $\eta \ll t_{\phi}^{-1}$, it is reasonable to neglect quantum correlations between sites of neighbouring chains, and treat the interchain coupling as a purely incoherent hopping process~\cite{jerome1982advphys}. This is expected to be a good approximation for temperatures intermediate between the two hopping energy scales, i.e.\  $\eta \ll k_B T \ll \tau$. This is easily satisfied in several systems. For example, $\eta\sim100$K and $\tau\sim1000$K for typical Bechgaard salts \cite{arrigoni2000prb}. In general, this separation of energy scales can occur for a number of reasons. For example, the interchain distance may be much larger than the separation between sites on the same chain. Alternatively, the hopping in the interchain direction might be mediated by a scaffolding of different molecules in between~\cite{dressel2005prb} (see figure~\ref{fig1}(b)).

By means of a Jordan-Wigner transformation, the present model can be mapped onto an interacting spinless fermion system \cite{Lieb1961aop}, as we demonstrate in \ref{JWtrans}. The parameter $\tau$ then corresponds to nearest-neighbour hopping, and $\tau\Delta$ to nearest-neighbour Coulomb repulsion, up to factors of order unity. Normally such a transformation is not feasible in a 2D system, due to the appearance of non-local Jordan-Wigner string operators (see equation~\eqref{JWstring}) which enforce the correct exchange phase between fermions at different sites. However, due to the purely incoherent nature of the interchain coupling, there is no need to maintain a definite phase relation between fermion states localised on different chains. This equivalence between fermion and spin representations makes our model relevant for describing not only spin transport, but also particle transport of hard-core bosons or fermions.

We describe the combination of coherent and incoherent dynamics by a quantum master equation of Lindblad form \cite{breuer}:
\begin{equation} \label{master_eq}
 \frac{\partial\rho}{\partial t}=-i[H,\rho] + \mathcal{L}(\rho),
\end{equation}
where $\rho$ is the density matrix of the total system, $H=\sum_{\lambda}H^{(\lambda)}$ is the total Hamiltonian, and $\mathcal{L}(\rho)$ is the dissipator describing the interaction of the spin chains with the environment and each other. The dissipator is given by
\begin{equation}
\mathcal{L}(\rho)=\sum_{k}\mathcal{L}_k(\rho)=\sum_{k}\biggl(L_{k}\rho L_{k}^{\dagger}-\frac{1}{2}\{L_{k}^{\dagger}L_{k},\rho\}\biggr),
\end{equation}
with $L_{k}$ the jump operators describing each incoherent process and $\{.,.\}$ the anticommutator of two operators. 

The incoherent coupling between two spin chains is modelled by the jump operators
\begin{equation}  \label{incoh_coup}
 L_{i}^{(\lambda,\mu)}=\sqrt{\gamma}\sigma_{i}^{+(\mu)}\sigma_{i}^{-(\lambda)}\delta_{\lambda,\mu\pm1},
\end{equation}
representing the transfer of a spin excitation from site $i$ of chain $\lambda$ to site $i$ of chain $\mu=\lambda\pm1$, with rate $\gamma$. Simple Golden Rule arguments \cite{jerome1982advphys} indicate that the incoherent hopping rate is of order $\gamma \sim \eta^2/t_{\phi}$. Due to the large number of factors that can contribute to this hopping rate (e.g.\ temperature, collision rate, interchain distance etc.), we treat $\gamma$ as a free parameter that can be varied independently. 

To analyse the transport properties of this system, we drive it into a non-equilibrium configuration by coupling its boundaries to unequal reservoirs, as depicted in figure \ref{fig1}(c). This driving scheme imposes a magnetisation imbalance on each chain, and thus induces a spin current. We assume that the correlation time of the reservoirs is negligibly small, so that the energy dependence of the incoherent transition rates may be neglected. We also assume that the points of contact between the bath and any pair of neighbouring chains are further apart than the correlation length \footnote{The correlation length is defined by $\lambda_c = c/\omega_c$, where $c$ is a characteristic velocity of bath excitations and $\omega_c$ is the frequency of the most energetic bath mode that the system interacts with appreciably. For example, for a fermionic bath consisting of a macroscopic metal lead the correlation length is essentially the Fermi wavelength, which gives $\lambda_c < 10^{-9}$ m for typical carrier concentrations greater than $10^{28}$ m$^{-3}$.}, leading to independent driving reservoirs for each chain. Under these conditions, it was shown in Ref.~\cite{benenti2009charge} that the reservoir degrees of freedom can be traced out under the standard Born-Markov approximation~\cite{breuer}. The action of the reservoirs driving the system out of equilibrium is therefore represented by the following Lindblad operators
\begin{equation} \label{lind_edges_coup}
L_{\text{L,R}}^{+(\lambda)}=\sqrt{\Gamma(1\pm f)/2}\sigma_{1,N}^{+(\lambda)}\qquad
L_{\text{L,R}}^{-(\lambda)}=\sqrt{\Gamma(1\mp f)/2}\sigma_{1,N}^{-(\lambda)}.
\end{equation}
Here, $\sigma^{\pm(\lambda)}_i=1/2(\sigma^{x(\lambda)}_i\pm i\sigma^{y(\lambda)}_i)$, $\Gamma$ is the strength of the coupling to the reservoirs (we choose $\Gamma=1$ in all our numerical calculations), and $f$ is the driving parameter. The driving operators are such that when applied to the boundary spins in isolation, they induce a state with magnetisation $\langle\sigma_1^{z(\lambda)}\rangle=f$ and $\langle\sigma_N^{z(\lambda)}\rangle=-f$. At $f=0$ there is no magnetisation imbalance between the boundaries of the chains, so there is no net spin transport. As $f$ increases, so does the imbalance between the boundaries, forcing a spin current to flow from the left to the right boundary of each chain. Equivalently, the Lindblad driving operators can be seen as injecting and ejecting spin excitations at different rates at each boundary, with $f$ determining the imbalance between these rates. Our simulations are performed with the weak driving $f=0.1$, thus staying in the linear response regime \cite{we,benenti2009charge}.

Due to the finite temperature, we would also expect local dephasing processes to exist, described by jump operators of the form
\begin{equation} \label{dephasing}
L_i^{(\lambda)}=\sqrt{\gamma_{\text{d}}/2}\sigma_i^{z(\lambda)},
\end{equation}
with $\gamma_{\text{d}}$ the dephasing rate. However, in order to simplify the analysis, in most of the paper we will assume that, apart from driving, the only effect of the environment is to generate incoherent hopping between chains. We show in Section \ref{mechanism} that the large current enhancement induced by incoherent coupling cannot result solely from dephasing processes. We also show in \ref{app_simultaneous} that our qualitative conclusions about the enhancement due to incoherent coupling remain valid in the simultaneous presence of dephasing.

\subsection{Mean-field approximation}

To gain insight into the properties of the system, we calculate its non-equilibrium steady state (NESS), which emerges in the long-time limit of equation~\eqref{master_eq} from the interplay between coherent and incoherent processes. Computing the NESS for a strongly correlated two-dimensional system represents a formidable challenge, therefore an approximation scheme is necessary. In \ref{analytic_approach} we present an exact solution for two incoherently coupled chains in the non-interacting limit $\Delta = 0$, demonstrating that the NESS factorises as $\rho = \rho_1 \otimes \rho_2 + O(f^2)$. Since magnetisation and current expectation values are of order $O(f)$, the lowest order contribution to the NESS is sufficient to compute transport observables accurately. This observation motivates the following mean-field approximation (MFA), according to which the state of the entire system is a direct product of the states of each spin chain
\begin{equation} \label{mean_field_ansatz}
\rho=\rho_1\otimes\rho_2\otimes\cdots\otimes\rho_{\Lambda},
\end{equation}  
thus discarding both quantum and classical correlations between different chains. Using this mean-field ansatz, we can obtain the master equation for each chain separately after tracing out the state of the other chains. This provides a considerable advantage in numerical simulations, which would be very demanding if all correlations between the chains are kept in the description. 

The resulting MFA master equation for chain $\lambda$ is

{\scriptsize
\begin{align} \label{master_eq_mf}
\frac{d\rho_{\lambda}}{dt}=&-i[H^{(\lambda)},\rho_{\lambda}]+\sum_{m=1}^N\tilde{\gamma}_m^{+(\lambda)}\biggl[\sigma_{m}^{+(\lambda)}\rho_{\lambda}\sigma_{m}^{-(\lambda)}-\frac{1}{2}\{\rho_{\lambda},\sigma_{m}^{-(\lambda)}\sigma_{m}^{+(\lambda)}\}\biggr]+\tilde{\gamma}_m^{-(\lambda)}\biggl[\sigma_{m}^{-(\lambda)}\rho_s\sigma_{m}^{+(\lambda)}-\frac{1}{2}\{\rho_{\lambda},\sigma_{m}^{+(\lambda)}\sigma_{m}^{-(\lambda)}\}\biggr],\nonumber\\
&\tilde{\gamma}_m^{+(\lambda)}=\frac{1}{2}\Gamma\Bigl[(1+f)\delta_{i,1}+(1-f)\delta_{i,N}\Bigr]+\frac{1}{2}\gamma[2+\langle\sigma_{m}^{z(\lambda+1)}\rangle+\langle\sigma_{m}^{z(\lambda-1)}\rangle]:=\tilde{\gamma}_m^{+\text{d}(\lambda)}+\tilde{\gamma}_m^{+\text{i}(\lambda)}\nonumber\\ 
&\tilde{\gamma}_m^{-(\lambda)}=\frac{1}{2}\Gamma\Bigl[(1-f)\delta_{i,1}+(1+f)\delta_{i,N}\Bigr]+\frac{1}{2}\gamma[2-\langle\sigma_{m}^{z(\lambda+1)}\rangle-\langle\sigma_{m}^{z(\lambda-1)}\rangle]:=\tilde{\gamma}_m^{-\text{d}(\lambda)}+\tilde{\gamma}_m^{-\text{i}(\lambda)},
\end{align}
}
\noindent
where the superscripts $\text{d}$ and $\text{i}$ refer to driving and incoherent coupling terms, respectively. Within the MFA, the incoherent interchain coupling turns into local gain and loss processes at each chain, with rates depending on the magnetisation of the neighbouring chains. In this form, the density matrix of each chain can be evolved separately from the others by the simulation of its own master equation, with the coupling to neighbouring chains being effectively described by expectation values of local operators. This type of evolution can be performed efficiently by means of a parallel implementation of the mixed-state Time Evolving Block Decimation (TEBD) algorithm \cite{zwolak2004mixed,cirac2004prl}. Our code is based on the open-source Tensor Network Theory (TNT) library \cite{tnt}.

\begin{figure}\centering
\includegraphics[scale = 0.3]{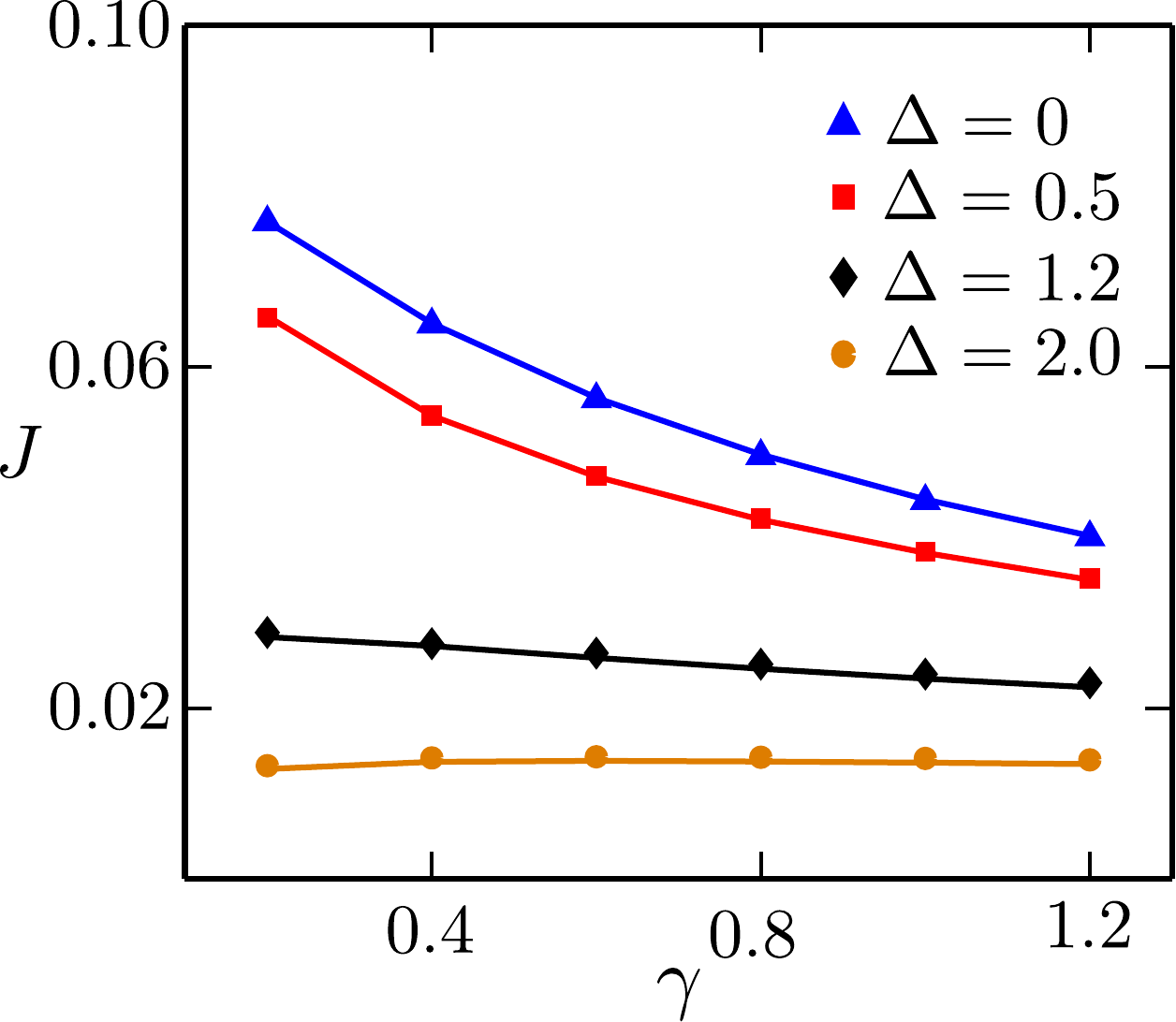}
\caption{Comparison between currents for two chains obtained with and without the MFA. Mean-field current expectation values are shown as symbols, and expectation values obtained without the MFA are shown as lines. \label{MF_comparisons}}
\end{figure}

In order to verify that the MFA gives reasonable results, we have also performed TEBD simulations of two coupled chains without the MFA for comparison. In figure\ \ref{MF_comparisons} we plot the steady-state currents (defined in Section \ref{current_definition}) obtained within each approach for a pair of incoherently coupled chains of length $N = 20$. The two sets of results are clearly in close agreement, however the accuracy of MFA calculations is higher for smaller values of $\gamma$ and $\Delta$. The maximum error is 3.8\%, when $\Delta = 2$ and $\gamma = 1.2$. We are therefore confident that this approximation gives accurate results for greater numbers of coupled chains, when quasi-exact TEBD simulations are not feasible.

\subsection{Approximation for an infinite number of coupled chains} \label{autocoupling}
In the case of an infinite number of chains, the reduced density operators of all the chains are exactly the same at any time. So as observed from equation~\eqref{master_eq_mf}, the problem of simulating the evolution of the entire system is reduced to that of performing the calculation for a single chain coupled twice with itself. The resulting Lindblad master equation of each chain, describing an effective non-linear self-consistent time evolution, is
\begin{equation} \label{master_eq_inf}
\frac{d\rho}{dt}=-i[H,\rho]+\sum_{m=1}^N\biggl[\tilde{\gamma}_m^{+}\Bigl(\sigma_{m}^{+}\rho\sigma_{m}^{-}-\frac{1}{2}\{\rho,\sigma_{m}^{-}\sigma_{m}^{+}\}\Bigr)+\tilde{\gamma}_m^{-}\Bigl(\sigma_{m}^{-}\rho\sigma_{m}^{+}-\frac{1}{2}\{\rho,\sigma_{m}^{+}\sigma_{m}^{-}\}\Bigr)\biggr],
\end{equation}
where the index $(\lambda)$ has been dropped for simplicity, and
\begin{equation} \label{effective_rates}
\tilde{\gamma}^{+}_m=\tilde{\gamma}_m^{+\text{d}}+\gamma(1+\langle\sigma^{z}_m\rangle)\qquad\tilde{\gamma}^{-}_m=\tilde{\gamma}_m^{-\text{d}}+\gamma(1-\langle\sigma^{z}_m\rangle)\\
\end{equation}
are the effective gain and decay rates at site $m$.

\subsection{Spin current} \label{current_definition}
We now derive the expression for the spin current through the system. It is obtained from the local magnetisation rate of change, calculated from the master equation directly. For site $i$ of chain $\lambda$, we have in the NESS
\begin{equation} \label{rate_of_change}
\biggl\langle\frac{d\sigma_i^{z(\lambda)}}{dt}\biggr\rangle=\text{Tr}\biggl(\sigma_i^{z(\lambda)}\frac{d\rho_{\lambda}}{dt}\biggr)=\langle j_{i-1}^{(\lambda)}\rangle-\langle j_{i}^{(\lambda)}\rangle
+\tilde{\gamma}_i^{+(\lambda)}(1-\langle\sigma_i^{z(\lambda)}\rangle)-\tilde{\gamma}_i^{-(\lambda)}(1+\langle\sigma_i^{z(\lambda)}\rangle)=0,
\end{equation}
where $j_i^{(\lambda)}$ is the longitudinal spin current through site $i$ of chain $\lambda$,
\begin{equation} \label{spin_curr_s}
j_i^{(\lambda)}=2(\sigma_i^{x(\lambda)}\sigma_{i+1}^{y(\lambda)}-\sigma_i^{y(\lambda)}\sigma_{i+1}^{x(\lambda)}).
\end{equation}
This expression is equivalent to that of the spin current through a 1D spin chain \cite{benenti2009charge}. Here, in contrast to that case, the longitudinal spin current is site-dependent in the NESS. For example, in the bulk of the system, $1<i<N$, the difference of spin currents through nearest neighbours is
\begin{equation} \label{diffe_currents_bulk}
\langle j_{i}^{(\lambda)}\rangle-\langle j_{i-1}^{(\lambda)}\rangle=\gamma\bigl(\langle\sigma_{i}^{z(\lambda+1)}\rangle+\langle\sigma_{i}^{z(\lambda-1)}\rangle-2\langle\sigma_i^{z(\lambda)}\rangle\bigr). 
\end{equation} 
\begin{figure}
\begin{center}
\includegraphics{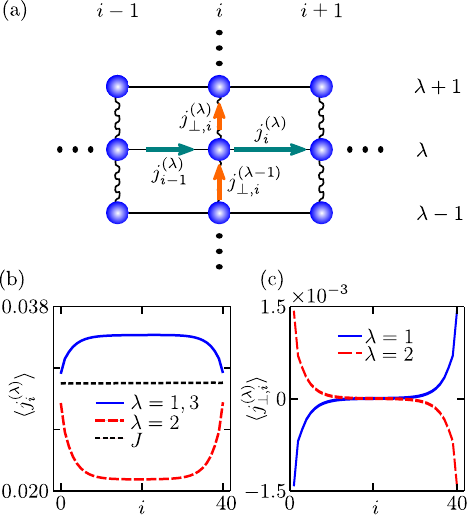}
\caption{\label{fig_continuity} (a) Diagram of three incoherently coupled spin chains, showing the different currents flowing through site $i$ of chain $\lambda$. The straight lines correspond to coherent coupling, while the curved lines represent incoherent coupling. (b). Longitudinal spin current for each chain of the system, and spin current per chain $J=\langle J_i\rangle/\Lambda$, with $\Lambda=3$, $N=40$, $\Delta=0.5$ and $\gamma=0.5$. (c). Transversal spin currents for the same parameters.}
\end{center}
\end{figure}
\noindent
Now consider, for example, the left boundary $i=1$. Since equation~\eqref{spin_curr_s} is not defined for $i=0$, in the NESS equation~\eqref{rate_of_change} gives
\begin{equation}
\langle j_{1}^{(\lambda)}\rangle=\gamma\bigl(\langle\sigma_{1}^{z(\lambda+1)}\rangle+\langle\sigma_{1}^{z(\lambda-1)}\rangle-2\langle\sigma_1^{z(\lambda)}\rangle\bigr)+\Gamma f-\Gamma\langle\sigma_1^{z(\lambda)}\rangle.
\end{equation}
A similar equation holds for the right boundary $i=N$. This leads to a natural definition of boundary currents $\langle j_0^{(\lambda)}\rangle$ and $\langle j_N^{(\lambda)}\rangle$, which allow equation~\eqref{diffe_currents_bulk} to be valid along the entire chain. These currents, which indicate the direct injection and ejection of spin excitations on the chain by the boundary reservoirs, are thus given by
\begin{equation} \label{currents_boundaries}
\langle j_0^{(\lambda)}\rangle=\Gamma f-\Gamma\langle\sigma_1^{z(\lambda)}\rangle\qquad\langle j_N^{(\lambda)}\rangle=\Gamma f+\Gamma\langle\sigma_N^{z(\lambda)}\rangle.
\end{equation}
\noindent
We can associate the right hand side of equation~\eqref{diffe_currents_bulk} to the difference of spin flows between chain $\lambda$ and its neighbouring chains, by defining a transversal spin current
\begin{equation}\label{transversal_curr}
\langle j_{\perp,i}^{(\lambda)}\rangle = -\gamma\bigl(\langle\sigma_i^{z(\lambda+1)}\rangle-\langle\sigma_i^{z(\lambda)}\rangle\bigr).
\end{equation}
In this form, equation~\eqref{diffe_currents_bulk} can be rewritten as
\begin{equation}
\langle j_{i-1}^{(\lambda)}\rangle+\langle j_{\perp,i}^{(\lambda-1)}\rangle=\langle j_{i}^{(\lambda)}\rangle+\langle j_{\perp,i}^{(\lambda)}\rangle.
\end{equation}
This balance between the longitudinal and transversal currents is illustrated in figure \ref{fig_continuity}(a).

From equation~\eqref{diffe_currents_bulk} it follows that in the absence of incoherent coupling, the current through each chain is homogeneous in the NESS, $\langle j_i^{(\lambda)}\rangle=\text{const}$, $i=0,\ldots,N$. In addition, from straightforward calculation, it is easily shown that in the presence of incoherent coupling, the total current per site $\langle J_i\rangle=\sum_{\lambda=1}^{\Lambda}\langle j_i^{(\lambda)}\rangle$ is homogeneous, i.e. that $\langle J_{i}\rangle-\langle J_{i-1}\rangle=0$. Also note that due to the symmetry of the system considered, $\sum_{\lambda=1}^{\Lambda-1}\langle j_{\perp,i}^{(\lambda)}\rangle=0$ for all sites $i$.

A concrete example of the longitudinal spin current profiles (equation~\eqref{spin_curr_s}) for three chains is shown in figure~\ref{fig_continuity}(b), including the boundary currents defined in equation~\eqref{currents_boundaries}. Due to the symmetry of the incoherent coupling, the state and thus the currents of chains $\lambda=1,3$ are equal. The corresponding transversal spin currents, defined in equation~\eqref{transversal_curr}, are shown in figure \ref{fig_continuity}(c). When moving from the boundary sites $i=1,N$ towards the centre of the system, the currents through chains $\lambda=1,3$ significantly increase at the expense of the current in the middle chain. This strong site dependence is reflected in large transversal currents, flowing in opposite directions. In the central sites of the system, the transversal currents are very small since the local magnetisations of neighbouring chains are very similar (see equation~\eqref{transversal_curr}). This is expected since the magnetisation of each chain must pass through zero at the same position in the centre due to the symmetric driving.

The NESS spin currents through the system, together with the magnetisation profile, determine the nature of the transport. If it is diffusive, the currents satisfy a diffusion equation:
\begin{equation} \label{diffusion_local}
\langle j_i^{(\lambda)}\rangle=-\kappa^{(\lambda)}\nabla\sigma_i^{z(\lambda)},
\end{equation}
where $\kappa^{(\lambda)}$ is the ($N$-independent) spin conductivity of chain $\lambda$ and 
\begin{equation} \label{magnet_grad}
\nabla\sigma_i^{z(\lambda)}=\langle\sigma_{i+1}^{z(\lambda)}\rangle-\langle\sigma_{i}^{z(\lambda)}\rangle
\end{equation}
is the magnetisation difference between neighboring spins of chain $\lambda$. On the other hand, if the transport is ballistic $\kappa^{(\lambda)}$ diverges, resulting in a size-independent spin current. Ballistic transport has been observed in single dephasing-free chains when $|\Delta|<1$ \cite{prosen2009matrix,benenti2009charge}, while diffusive transport has been found in the linear response regime for $|\Delta|>1$ and no dephasing \cite{prosen2009matrix,benenti2009charge}, and for finite dephasing and any interaction strength \cite{znidaric2010exact,znidaric2011solvable,znidaric2010dephasing}. Note that since the transversal current of equation~\eqref{transversal_curr} is proportional to the local magnetisation difference along the transversal direction, it is diffusive by construction.

We now discuss the spin transport properties of the system in both the weakly ($|\Delta|<1$) and strongly ($|\Delta|>1$) interacting regimes, which show a completely different behaviour in the presence of incoherent interchain coupling. For this, instead of observing the spin current through each chain, we consider the total spin current \textit{per chain}, noted by $J$, i.e., $J=\langle J_i\rangle/\Lambda$. Thus $J\Lambda$ is the total spin current per site in the NESS. We refer to $J$ in the rest of the paper simply as the spin current; it is shown in figure \ref{fig_continuity}(b) for a particular example. Its homogeneity along the system is a good indication of the obtention of the NESS.

\section{Transport in weakly interacting incoherently coupled spin chains} \label{weak}

\begin{figure}
\begin{center}
\includegraphics{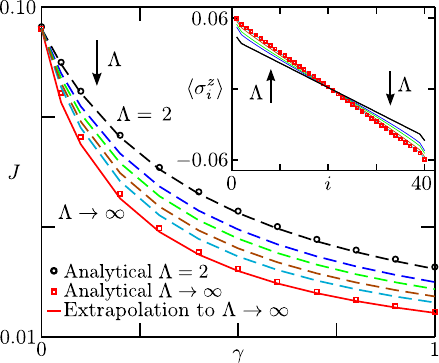}
\caption{\label{D0_Lvar} Spin current as a function of incoherent coupling $\gamma$ for several numbers of non-interacting ($\Delta=0$) chains of $N=40$. The arrow indicates the decreasing tendency of the current as $\Lambda$ increases. The dashed lines correspond to results of TEBD simulations ($\Lambda=2,3,4,6,10$). The red solid line indicates an extrapolation of results of a finite number of chains (up to $\Lambda=10$) to $\Lambda\rightarrow\infty$, using a simple rational function. The symbols indicate the analytical calculation for $\Lambda=2$ ($\circ$) and $\Lambda\rightarrow\infty$ ($\scriptstyle\square$). Inset: Magnetisation profile of a chain in the centre of the system ($\Lambda/2$ for $\Lambda$ even, $(\Lambda+1)/2$ for $\Lambda$ odd) for $\gamma=0.3$ and the same number of chains shown in the main panel. The solid lines correspond to the magnetisations obtained from TEBD results, and the symbols ($\scriptstyle\square$) to the analytical approach for the self-coupled chain.}
\end{center}
\end{figure}

We initially consider the non-interacting case $\Delta=0$, which leads to the same nearest-neighbour coherent coupling as is frequently considered in toy models of exciton transport in light harvesting complexes~\cite{deph_assisted_plenio,chin2010njp}. The analytical method presented in Refs.~\cite{znidaric2010exact,znidaric2011solvable} can be extended to two incoherently coupled non-interacting chains, as explained in \ref{analytic_approach}. This allows us to extract the exact current and magnetisation expectation values. The former is given by
\begin{equation} \label{curr_analytic_L2}
J = \frac{4f}{(\Gamma/4)+(4/\Gamma)+(N-1)\gamma/4},
\end{equation}
while the magnetisation profile is linear in the bulk, see equation \eqref{magnetisation_d0}. These results agree with TEBD simulations, as indicated in figure \ref{D0_Lvar}. Note that if $\gamma=0$, the current is independent of the size of the spin chains, indicating ballistic transport \cite{znidaric2010exact,znidaric2011solvable}. On the other hand, a finite incoherent coupling induces a decay of the current with the length of the system $\propto N^{-1}$, typical of a diffusive conductor. In fact the bulk conductivities are easily shown to be $\kappa^{(1,2)}=8/\gamma$. So, similarly to dephasing processes on a single non-interacting chain \cite{znidaric2010exact,znidaric2011solvable}, incoherent interchain couplings induce a non-equilibrium phase transition between ballistic and diffusive regimes, with a spin current monotonically degraded by the interchain hopping.

In the limit of an infinite number of chains, described by the self-coupled chain (see Section \ref{autocoupling}), the analytical method used for two chains can also be applied (see \ref{analytic_approach}). We thus obtain exact expressions for the current and magnetisation, given by equations \eqref{curr_analytic_L2} and \eqref{magnetisation_d0} respectively, with $\gamma/2$ instead of $\gamma/4$. The conductivity is thus $\kappa=4/\gamma$, reduced compared to the case of $\Lambda=2$. This is because each chain has two nearest neighbours rather than one, leading to a stronger degrading effect of the incoherent hopping.

In the intermediate case, i.e. for a finite number of chains $\Lambda>2$, we use the TEBD method to obtain the NESS of the system. Characteristic results for the current and for the magnetisation profiles are shown in figure~\ref{D0_Lvar}. The same qualitative features of the cases $\Lambda=2$ and $\Lambda\to\infty$ are found, namely the spin current monotonically decreases with $\gamma$ and the magnetisation profile is almost linear in the bulk. In addition, for a fixed incoherent coupling, the current decreases with the number of chains, rapidly for small values of $\Lambda$ and very slowly for large values. An extrapolation of these results to the limit $\Lambda\to\infty$ agrees with the analytical approach for the self-coupled chain, as shown in figure~\ref{D0_Lvar}.

\begin{figure}
\begin{center}
\includegraphics{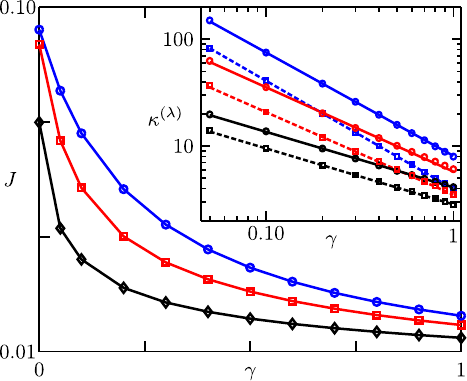}
\caption{\label{Current_Dless1} Spin current as a function of $\gamma$ for $\Lambda=10$, $N=40$ and different interaction strengths $\Delta<1$. The symbols correspond to TEBD results: blue ($\circ$) to $\Delta=0$, red ($\scriptstyle\square$) to $\Delta=0.5$ and black ($\diamond$) to $\Delta=0.9$; the solid lines are guides to the eye. The qualitative behaviour of the current, i.e. its monotonic decay with $\gamma$, is observed for any other value of $\Lambda$. Inset: Conductivities for $\lambda=1$ ($\circ$, solid lines), indicating the behaviour at the boundary chains of the system, and of $\lambda=2$ ($\scriptstyle\square$, dashed lines), corresponding to the bulk. The symbols are TEBD results, and the lines are fits to the power law $\kappa^{(\lambda)}=\alpha_{\lambda}\gamma^{-\beta_{\lambda}}$. For $\Delta=0$, $\beta_1=0.968(4)$ and $\beta_2=1.013(2)$. For $\Delta=0.5$, $\beta_1=0.79(2)$ and $\beta_2=0.77(2)$. For $\Delta=0.9$, $\beta_{1,2}=0.52(1)$. The colors correspond to the same interaction strengths $\Delta$ of the main panel.}
\end{center}
\end{figure}

Similarly to the cases of two and an infinite number of chains, the transport response induced by incoherent coupling on a system of several chains is characteristic of diffusive conductors. To see this explicitly, we observe that each chain satisfies the local diffusion equation~\eqref{diffusion_local}. Since the spin current through each chain is site-dependent, we have verified that the ratio $\langle j_i^{(\lambda)}\rangle/\nabla\sigma_i^{z(\lambda)}$ (i.e. the conductivity) is homogeneous for each $\lambda$, so the diffusion equation~\eqref{diffusion_local} holds. In addition, similarly to the analytically-solvable cases, the conductivity of each chain decays monotonically with the incoherent coupling rate, with a behaviour very well described by a decay $\kappa^{(\lambda)}\propto1/\gamma$, as shown in figure \ref{Current_Dless1}. We also note that for $2<\lambda<\Lambda-1$ the conductivity is almost indistinguishable from that of $\lambda=2,\Lambda-1$, due to the weak effect of the boundary chains.

Now we consider weak interactions $0<\Delta<1$. We find that the effect of incoherent interchain coupling on the system is very similar to that on non-interacting chains. Namely a finite incoherent coupling induces a transition from ballistic ($\gamma=0$) to diffusive ($\gamma>0$) behaviour. The magnetisation profiles become linear, and the spin current and the conductivities of each chain decrease monotonically with $\gamma$, the latter following a power law as shown in figure \ref{Current_Dless1}. The current also decreases with $\Lambda$, approaching a limiting value when $\Lambda\rightarrow\infty$. In figure \ref{Current_Dless1} it is also seen that for fixed values of $\Lambda$ and $\gamma$, the spin transport diminishes as $\Delta$ increases, a known result for single chains in the massless regime \cite{gobert2005real,znidaric2010dephasing}.

\begin{figure}
\begin{center}
\includegraphics{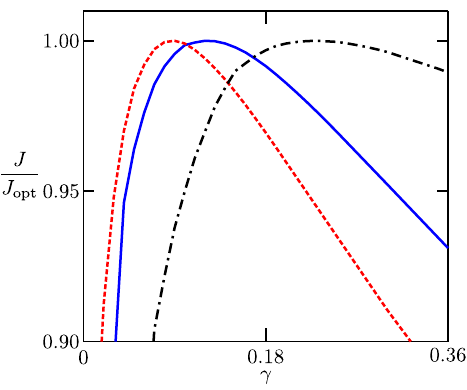}
\caption{\label{assistance_incoh} Spin transport enhancement in the strongly interacting regime due to incoherent interchain hopping. For clarity we plot the rescaled spin current $J/J_{\text{opt}}$ as a function of $\gamma$. The solid blue line corresponds to $\Delta=1.2$, $\Lambda=3$ and $N=40$. The other lines indicate the effect on the spin current when some parameters of the system are modified, i.e. when the number of chains increases (dashed red line, $\Delta=1.2$ and $\Lambda=10$), and when the interaction is stronger (dot-dashed black line, $\Delta=2$ and $\Lambda=3$). When $\gamma=0$, $J=4.25\times10^{-3}$ for $\Delta=2$, and $J=1.19\times10^{-2}$ for $\Delta=1.2$. The corresponding optimal currents are $J_{\text{opt}}=8.0\times10^{-3}$ and $J_{\text{opt}}=1.65\times10^{-2}$ respectively.}
\end{center}
\end{figure}

\section{Transport enhancement for strong intrachain coupling} \label{subassisted}
We now consider the effect of incoherent interchain coupling on the transport properties of strongly interacting spin chains ($\Delta>1$).

\subsection{Environment assisted transport}
Due to the strong correlations between spin excitations, the regime $\Delta>1$ presents a completely different response to environmental effects to the case of weak interactions $\Delta<1$. It has been found \cite{we} that for single 1D chains, dephasing processes can lead to a surprisingly large enhancement of the current even at weak driving. Now we show that the ability of excitations to jump incoherently across different chains leads to an even larger transport enhancement, which constitutes the main result of our work. 

As shown in figure~\ref{assistance_incoh}, the presence of incoherent interchain coupling increases the spin current through the system, compared to that of $\gamma=0$, for a wide range of rates $\gamma$. The optimal coupling maximizing the current $\gamma_{\text{opt}}$, which is obtained by fitting the peak to a polynomial function, strongly depends on the interaction strength, increasing with $\Delta$. Similarly, the current enhancement grows with $\Delta$. For example, the current increases by up to $39\%$ for $\Delta=1.2$ and up to $91\%$ for $\Delta=2$. 

\begin{figure}
\begin{center}
\includegraphics{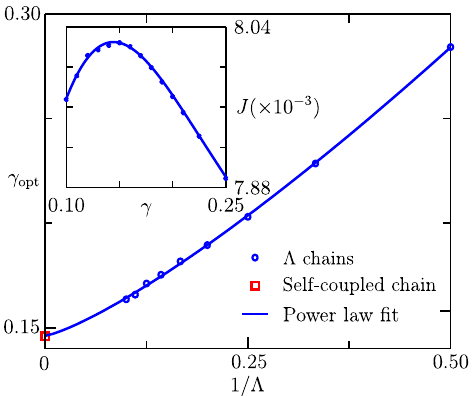}
\caption{\label{fig10chains} Optimal incoherent coupling as a function of the number of coupled chains $\Lambda$, for $N=40$ and $\Delta=2$ ($\circ$), and the optimal coupling for a self-coupled chain (\scalebox{0.6}{$\square$}). The power law fit $\gamma_{\text{opt}}=a\Lambda^{-b}+\gamma_{\text{opt}}^{\infty}$ is shown (solid line), which results in an optimal coupling at $\Lambda\rightarrow\infty$ of $\gamma_{\text{opt}}^{\infty}=0.146(5)$; $a=0.33(1)$, $b = 1.3(1)$. Inset: simulations of the self-coupled chain provide an alternative method for finding the optimal coupling in the limit $\Lambda\to\infty$. Fitting the results of the simulations around the peak ($\circ$) to a polynomial function (solid line), we find the maximum current at $\gamma_{\text{opt}}^{\infty} = 0.144$, consistent with the scaling analysis.}
\end{center}
\end{figure}

We now consider the effect of the system size on the transport enhancement. The optimal current $J_{\text{opt}}$ remains almost constant for all values of $\Lambda$ considered. In addition, the optimal coupling monotonically decreases with $\Lambda$ as shown in figure~\ref{fig10chains}, and the range of beneficial couplings narrows. Importantly, extrapolations to $\Lambda\rightarrow\infty$ strongly suggest the existence of a finite optimal incoherent coupling $\gamma_{\text{opt}}^{\infty}$ in this limit. We have confirmed this result from simulations of a self-coupled chain with $\Delta=2$, as shown in the inset of figure~\ref{fig10chains}. The results of both approaches agree very well, giving optimal couplings of $\gamma_{\text{opt}}^{\infty}=0.146$ for the extrapolation and $\gamma_{\text{opt}}=0.144$ for the self-coupled chain. Because of this agreement, we henceforth denote by $\gamma_{\text{opt}}^{\infty}$ the optimal coupling for the self-coupled chain. 

Note that a simple argument qualitatively explains the decay of $\gamma_{\text{opt}}$ with $\Lambda$. Consider first the case $\Lambda=2$. Since each chain is only affected by just a single neighbouring chain, it is expected that  $\gamma_{\text{opt}}(\Lambda=2)\sim2\gamma_{\text{opt}}^{\infty}$, as seen in figure~\ref{fig10chains}. For $\Lambda=3$, the boundary chains are coupled to a single neighbour, so their transport is optimised by $\gamma\sim2\gamma_{\text{opt}}^{\infty}$. The central chain, being coupled to two neighbours, is optimised by $\gamma\sim\gamma_{\text{opt}}^{\infty}$. Assuming that an average incoherent coupling optimises the transport of the entire system, we get $\gamma_{\text{opt}}(\Lambda=3)\sim5\gamma_{\text{opt}}^{\infty}/3$. In general, for $\Lambda$ chains, we expect
\begin{equation}
\gamma_{\text{opt}}(\Lambda)\sim\frac{(\Lambda-2)\gamma_{\text{opt}}^{\infty} + 4\gamma_{\text{opt}}^{\infty}}{\Lambda}=\frac{2\gamma_{\text{opt}}^{\infty}}{\Lambda}+\gamma_{\text{opt}}^{\infty}.
\end{equation}
This simple scaling provides a good approximation to that found from the power law fit of the results of a finite number of chains, as indicated in figure~\ref{fig10chains}.

\begin{figure}
\centering
\includegraphics{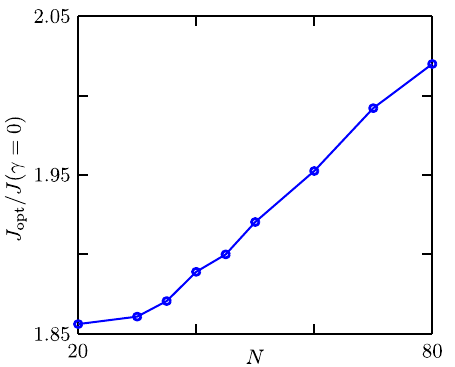}
\caption{Spin current enhancement factor $J_{\text{opt}}/J(\gamma=0)$ as a function of the length of the system for a self-coupled chain with $\Delta=2$ ($\circ$). The solid line is a guide to the eye. \label{scaling_enhancement}}
\end{figure}

To observe how the enhancement effect scales with the length of the system, we have analysed both the optimal current and the optimal coupling for self-coupled chains with different values of $N$. We found that an exponential decay of the optimal coupling of the form $\gamma_{\text{opt}}^{\infty} = a e^{-b N} + c$ yields a finite optimal coupling in the thermodynamic limit of $\gamma_{\text{opt}}^{\infty}(N\to\infty) = c = 0.057(9)$. Nevertheless, a power law decay of the form $\gamma_{\text{opt}}^{\infty} =aN^{-b}$ also fits well to our results. This means that we are not able to assess whether the optimal coupling is finite for an infinite system. The scaling results indicate, however, that for very large but finite systems, an enhancement effect of the current is still expected for very small incoherent couplings. This does not mean that the increase of the current becomes less important as the system gets larger. In fact, although it is restricted to a narrower range on incoherent coupling rates, the enhancement effect becomes stronger as the size of the system increases. This is shown in figure~\ref{scaling_enhancement}, where the enhancement factor $J_{\text{opt}}/J(\gamma=0)$ is seen to increase with $N$. We therefore expect that spin transport can be significantly enhanced by environmental processes even in macroscopic (anisotropic) two-dimensional systems, and thus can be observed experimentally in bulk materials.

\begin{figure}
\centering
\includegraphics{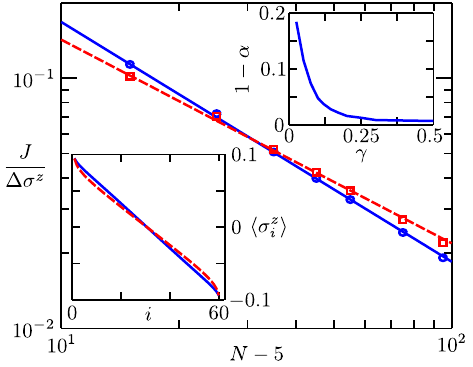}
\caption{Scaling of $J/\Delta\sigma^z$ of the self-coupled chain, for $\Delta=2$, $\gamma=0.05$ ($\scriptstyle\square$) and $\gamma=0.20$ ($\circ$). The results of the simulations fit very well to equation~\eqref{fit_self_coupled}, giving $\kappa=0.93(6)$ and $\alpha=0.81(2)$ for $\gamma=0.05$ (dashed red line), and $\kappa=1.53(3)$ and $\alpha=0.959(6)$ for $\gamma=0.20$ (solid blue line). Lower inset: Corresponding magnetisation profiles of $N=60$. Upper inset: $1-\alpha$ as a function of $\gamma$. \label{J_scaling_selfCoupled}}
\end{figure}

To understand the origin of the increase of the enhancement ratio with $N$, it is important to study the nature of the spin transport in the enhancement regime. To address this point we analyse the scaling with $N$ of the spin current through a strongly interacting self-coupled chain. The results are shown in figure~\ref{J_scaling_selfCoupled}. In the presence of diffusive spin transport, the magnetisation profile of the chain is linear in the bulk. The local magnetisation difference is thus homogeneous, and defined as 

\begin{equation}
\nabla\sigma_i^z=\frac{\Delta\sigma^z}{N-5}\quad\text{with}\quad\Delta\sigma^z =\langle\sigma_{N-2}^z\rangle-\langle\sigma_3^z\rangle,
\end{equation}
where $N-5$ corresponds to removing two sites from either end to diminish boundary effects. Diffusive spin transport is evidenced if the system satisfies the diffusion equation
\begin{equation} \label{fit_self_coupled}
\frac{J}{\Delta\sigma^z}=-\frac{\kappa}{(N-5)^{\alpha}},
\end{equation}
with $\kappa$ the spin conductivity and $\alpha=1$. As shown in figure\ \ref{J_scaling_selfCoupled} for weak incoherent coupling, the results of our TEBD simulations are well described by this equation, but with $\alpha<1$. This means that in the regime of transport enhancement, the system presents a spin current which decreases with $N$ slower than normal diffusion, i.e. it shows superdiffusive behaviour. Thus the optimal current also shows a slower decrease than that of a diffusive conductor. Since for the diffusive regime $J(\gamma=0)\propto N^{-1}$, a divergence of the enhancement ratio with $N$ is found. 

Finally, as shown in the upper inset of figure\ \ref{J_scaling_selfCoupled}, the exponent $\alpha$ gets closer to 1 when increasing $\gamma$, the transport thus tending towards being described by normal diffusion when the incoherent effects become stronger. When $\gamma$ is too large the enhancement effect disappears, since the system is perturbed so frequently that it is prevented from evolving, i.e.\ the Zeno effect emerges \cite{breuer}.

\subsection{Enhancement mechanism} \label{mechanism}

\begin{figure}
\centering
\includegraphics[scale=0.8]{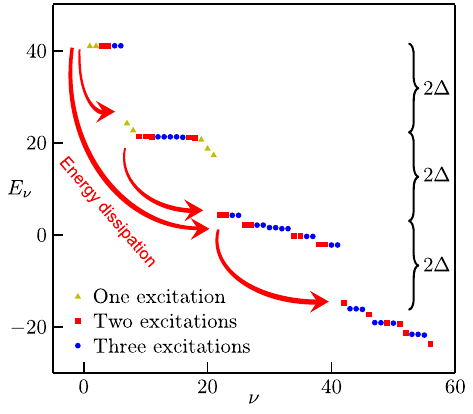}
\caption{Cartoon of the band structure of the $XXZ$ Hamiltonian in the strongly interacting regime. The highest bands consist of bound states with low conductivity, while lower bands contain more mobile states. The red arrows indicate possible transitions induced by energy dissipation, between states of the same or different spin sectors. The energy values were obtained from a chain of $N=7$ and $\Delta=10$. \label{bandStructure}}
\end{figure}

We now discuss the origin of the transport enhancement in the strongly interacting regime, which is similar to that found in single chains due to dephasing processes \cite{we,we2}. For interaction strengths $\Delta>1$, the spectrum of the $XXZ$ Hamiltonian~\eqref{hami} consists of several bands whose energetic separation is proportional to $\Delta$ (see figure \ref{bandStructure}). The highest bands are almost flat, possessing very low conductivity. These bands correspond to bound states of spin excitations, where several spins are clumped together, thus having large potential energy.

When the system is driven out of equilibrium, population is transferred to various eigenstates, depending on the strength of the driving. For example, at $f = 1$ only the highest bands are populated, leading the system to an insulating NESS \cite{we,benenti2009charge}. Even in the weak-driving regime as considered here, some population is transferred to the highest bands, which then gives a small contribution to the conduction of the system. However, if energy-dissipating processes take place in the system, transitions from these slow bands to lower bands of larger conductivity are induced, leading to an enhancement of the current. In other words, if the energy of spin bound states is dissipated, these break into states of lower potential (and total) energy, but of much higher kinetic energy, thus increasing the conductivity. 

The enhancement described in our work emerges from the energy dissipation induced by the incoherent interchain coupling. To clarify this point, consider for simplicity the self-coupled chain configuration described by equation~\eqref{master_eq_inf}  \footnote{The calculation of the rate of energy dissipation is also easily performed for a finite number of chains. In this case, we obtain a sum over $\lambda$ of terms like those of equation~\eqref{dissipation_self}, but instead of the magnetisation at site $i+1$ of each chain, the magnetisations of the neighbouring chains at site $i+1$ appear. Nevertheless, since the magnetisations of all chains are similar, equation~\eqref{dissipation_self} corresponds to a good approximation.}. A straightforward calculation of the energy dissipation rate corresponding to the incoherent coupling gives
\begin{equation}\label{dissipation_self}
\text{Tr}(H\mathcal{L}_{\text{inc}}(\rho))=-2\gamma\sum_{j=1}^{N-1}\langle\sigma_i^x\sigma_{i+1}^x+\sigma_i^y\sigma_{i+1}^y\rangle-4\gamma\Delta\sum_{j=1}^{N-1}(\langle\sigma_i^z\sigma_{i+1}^z\rangle-\langle\sigma_i^z\rangle\langle\sigma_{i+1}^z\rangle),
\end{equation}
where $\mathcal{L}_{\text{inc}}(\rho)$ is the dissipator describing the incoherent coupling. The first term in the energy dissipation rate appears due to the loss of phase coherence between neighbouring sites \cite{we}, and is proportional to the hopping energy
\[K = \sum_{i=1}^{N-1} \langle\sigma_i^x\sigma_{i+1}^x+\sigma_i^y\sigma_{i+1}^y\rangle. \]
The second term is proportional to the sum of nearest-neighbour connected spin-spin correlations
\[C = \sum\limits_{i=1}^{N-1} (\langle \sigma^z_i \sigma^z_{i+1} \rangle - \langle\sigma^z_i\rangle\langle\sigma^z_{i+1}\rangle)  ,\]
and corresponds to a direct dissipation of the interaction energy due to the incoherent hopping, which rips spin excitations away from their nearest neighbours \footnote{As observed during the derivation of equation~\eqref{dissipation_self}, the terms $\protect\langle\sigma_i^z\sigma_{i+1}^z\protect\rangle$ are a direct consequence of the non-conserving nature of the jump operators (i.e. they result from any incoherent process described by jump operators $\sigma^+$ or $\sigma^-$). In addition, the terms $\protect\langle\sigma_i^z\protect\rangle\protect\langle\sigma_{i+1}^z\protect\rangle$ appear because the effective rates $\tilde{\gamma}^{+}_m$ and $\tilde{\gamma}^{-}_m$ are different (see equation~\eqref{effective_rates}).}.

\begin{figure}
\centering
\includegraphics{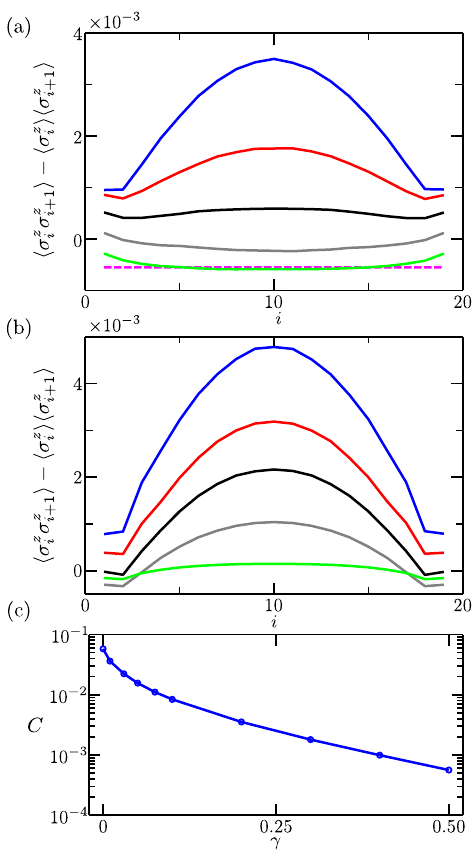}
\caption{Nearest-neighbour connected correlation function for different parameters of the self-coupled chain of $N=20$. (a) For $\gamma=0$ and several interaction strengths $\Delta$. From bottom to top, the solid lines correspond to $\Delta=0.8,1.0,1.05,1.1,1.2$. The dashed line refers to $\Delta=0$. (b) For $\Delta=1.4$ and several incoherent coupling rates $\gamma$. From top to bottom, the lines correspond to $\gamma=0,0.01,0.03,0.1,0.5$. (c) Sum of the nearest-neighbour correlation functions $C$ as a function of $\gamma$ for $\Delta=1.4$. \label{correlationFunctions}}
\end{figure}

Intuitively, if $C$ is positive, the spin excitations of the system are bunched together on average, while if $C < 0$ the excitations are spread out. Therefore, the sign of $C$ gives a simple indication of how population is distributed between the bound states (bunched) and mobile states (spread out). In the absence of incoherent coupling, we have found that $C$ undergoes a marked transition from taking negative to positive values as the interaction strength crosses the critical point $\Delta = 1$ (see figure~\ref{correlationFunctions}(a)). This behaviour is a manifestation of the well known non-equilibrium phase transition from ballistic to diffusive conduction at $\Delta = 1$ \cite{we,benenti2009charge}, and demonstrates a tendency of the spin excitations to clump together in the strongly interacting regime of the driven system. Importantly, even when the incoherent coupling is incorporated, our simulations always show that $C>0$ when $\Delta > 1$ (see figures~\ref{correlationFunctions}(b),(c)). More precisely, $C$ diminishes with $\gamma$, indicating the decrease of population in bound (correlated) states with incoherent processes, but remains positive. Similarly, we always found that $K>0$ in the strongly interacting regime. This indicates that energy is being dissipated from the system due to the incoherent interchain hopping (see equation \eqref{dissipation_self}), transferring population from bound to mobile states and thus leading to transport enhancement.

\begin{figure}
\begin{center}
\includegraphics{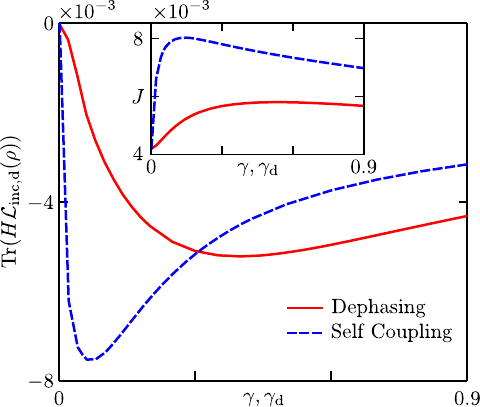}
\caption{\label{comparison_self_deph} Comparison of the rate of energy dissipation due to bulk dephasing processes (equation~\eqref{dissipation_dephasing}), and incoherent self coupling (equation~\eqref{dissipation_self}). Both cases correspond to $\Delta=2$ and $N=40$. For these parameters, the current for a single isolated chain is $J\approx4.2\times10^{-3}$. Inset: Corresponding spin currents.}
\end{center}
\end{figure}
 
It is important to note that the transport enhancement induced by incoherent interchain coupling is much larger than that of pure dephasing processes described by equation~\eqref{dephasing}. For example, as shown in figure\ \ref{comparison_self_deph} for $\Delta=2$, the spin current is increased up to $37\%$ by dephasing \cite{we}, and up to $91\%$ by incoherent coupling. This difference can be explained by looking at the energy dissipation rate due to dephasing,
\begin{equation}\label{dissipation_dephasing}
\text{Tr}(H\mathcal{L}_{\text{d}}(\rho))=-2\gamma_{\text{d}}\sum_{j=1}^{N-1}\langle\sigma_i^x\sigma_{i+1}^x+\sigma_i^y\sigma_{i+1}^y\rangle,
\end{equation}
with $\mathcal{L}_{\text{d}}(\rho)$ the dephasing dissipator. This rate is compared to the dissipation rate from incoherent coupling in figure\ \ref{comparison_self_deph}. Since its maximal magnitude is significantly smaller than that of incoherent coupling, more energy is dissipated by the latter, resulting in more population transfer from flat to mobile energy bands and thus to a larger current. This also shows that for $\Delta\neq0$, the effects of incoherent coupling cannot be reproduced just by dephasing processes. In \ref{app_simultaneous} we also show that in the simultaneous presence of dephasing and incoherent coupling, the latter dominates the energy dissipation, and the current enhancement is still larger than that of dephasing alone. 

\section{Summary \& Conclusions} \label{conclu}

We have studied the spin transport in an anisotropic 2D spin$-\frac{1}{2}$ lattice driven out of equilibrium by Markovian boundary reservoirs. The assumption of highly anisotropic coupling allowed us to consider the system as an array of incoherently coupled chains. Each chain is described by an $XXZ$ Hamiltonian, which contains the basic elements that constitute many-body lattice systems, namely particle hopping and interactions. Employing a mean-field approximation, we calculated the spin current and magnetisation of the NESS of the system for several parameters. This approximation, found to reproduce the transport properties of two coupled chains, facilitates an accurate and efficient dynamical simulation of the system using a parallel implementation of the TEBD algorithm \cite{tnt}. 

We found that in the presence of weak intrachain interactions, the incoherent coupling monotonically degrades the spin conductivity of the chains. However, in the strongly interacting regime we found a significant transport enhancement due to the incoherent coupling. This enhancement can be understood as the result of incoherent transitions from bound states to mobile bands of energy eigenstates, similar to the effect of dephasing on spin and heat transport in 1D systems \cite{we,we2}. However, the direct breakdown of bound states by the incoherent hopping between neighbouring chains, which opens more paths for spin excitations to flow, provides a greater improvement than dephasing effects alone. 

A self-consistent extension of the mean-field approximation enabled us to perform simulations directly in the limit of an infinite number of chains. In this configuration, we found that the enhancement of the spin current increases with the size of the system, which reveals the importance that the enhancement effect can have in bulk materials. The origin of this scaling was related to the existence of superdiffusive transport in the regime of current enhancement, becoming closer to normal diffusion as the incoherent coupling increases.

Finally, we note that the effects described in our work has so far not been found experimentally. Real materials such as organic conductors and cuprates involve more complicated effects than those considered here, which would have a significant impact on their transport properties, and may obstruct a direct observation of the enhancement and degradation mechanisms we have described. In particular, in the absence of interactions, we expect that incoherent interchain hopping destroys disorder-induced localisation. However, when interactions are added to the picture, it is not clear how the two transport enhancement effects would combine. A model incorporating disordered site energies alongside interactions and incoherent hopping would be amenable to a numerical study using the methods described in this work. This constitutes an interesting topic for future research. Alternatively, quantum simulators such as ultracold atom systems are intrinsically free of disorder. Moreover, several experimental \cite{brantut2012sci, stadler2012nat} and theoretical \cite{caruso2011pra,bermudez2013prl} advances aimed at simulating quantum transport in such systems have recently been made. This offers the prospect of observing the effects described in this work and studying them in the laboratory using current or near-future technology.

\ack
The research leading to these results has received funding from the European Research Council under the European Union's Seventh Framework Programme (FP7/2007-2013) / ERC Grant Agreement n$^{\mathrm{o}}$ 319286, and from the EPSRC projects EP/K038311/1 and EP/J010529/1. We acknowledge Sarah Al-Assam, Chris Goodyer and the TNT Library Development Team for providing the codes for the simulations carried out during our work. J.J.M.-A. acknowledges Departamento Administrativo de Ciencia, Tecnolog\'{i}a e Innovaci\'{o}n Colciencias for economic support, and Thomas Grujic for his help at the beginning of this project. M.T.M. acknowledges support from the UK EPSRC via the Controlled Quantum Dynamics CDT. D.J. and S.R.C. thank the National Research Foundation and the Ministry of Education of Singapore for support.  J.P. thanks the Spanish Ministerio de Econom\'{i}a y Competitividad under Project No. FIS2012-30625. M.B.P. acknowledges support from the Alexander von Humboldt Foundation, the ERC Synergy grant BioQ and the EU STREP project PAPETS. 

\appendix

\section{Equivalence of fermion and spin representations for incoherently coupled chains}
\label{JWtrans}

We start from a fermionic tight-binding model, describing $\Lambda$ incoherently coupled 1D chains of $N$ sites each. Using the Jordan-Wigner transformation, we can map this system onto the spin model described in the main text. We consider only the incoherent interchain coupling, since the Jordan-Wigner mapping for the boundary driving Lindblad operators is detailed in Ref. \cite{benenti2009charge}. To carry out the proof, it is simplest to work in the Heisenberg picture. The evolution of an operator $O$ is given by
\begin{equation}
\label{heisenbergEvolution}
\frac{\partial O}{\partial t} = i \sum\limits_{\lambda = 1}^{\Lambda} [ H_F^{(\lambda)}, O] + \sum\limits_{\lambda=1}^{\Lambda-1}\sum\limits_{j=1}^{N} \mathcal{A}_{j}^{(\lambda) \dagger}(O).
\end{equation}
The Hamiltonian of each chain is 
\begin{equation}
\label{fermionH}
H_F^{(\lambda)} = -2\tau \sum\limits_{j=1}^{N-1} \left ( c_{j}^{(\lambda) \dagger}c^{(\lambda)}_{j+1} + c_{j+1}^{(\lambda) \dagger}c^{(\lambda)}_{j}-2\Delta n^{(\lambda)}_{j}n^{(\lambda)}_{j+1}\right ) ,
\end{equation}	
where the ladder operators satisfy $\{c^{(\lambda)}_{j}, c^{(\mu) \dagger}_{k}\} = \delta_{j,k}\delta_{\lambda,\mu}$, and $n_j^{(\lambda)} = c^{(\lambda) \dagger}_{j} c^{(\lambda)}_{j}$. The adjoint dissipator describing the hopping between sites $j$ of chains $\lambda$ and $\lambda+1$ is 
\begin{equation}
\label{fermionSiteLiouvillian}
\mathcal{A}_{j}^{(\lambda) \dagger}(O)\; = \; A_{j}^{(\lambda)} O A_{j}^{(\lambda) \dagger} - \frac{1}{2} \{ A_{j}^{(\lambda)} A_{j}^{(\lambda) \dagger}, O \} +  A_{j}^{(\lambda) \dagger} O A_{j}^{(\lambda)} - \frac{1}{2} \{ A_{j}^{(\lambda) \dagger} A_{j}^{(\lambda)}, O \} ,
\end{equation}
where 
\begin{equation}
\label{fermionJumpOp}
A_{j}^{(\lambda)} = \sqrt{\gamma} c_{j}^{(\lambda + 1) \dagger} c_{j}^{(\lambda)}.
\end{equation}

Each site of the system is associated to an index pair $(j,\lambda)$ specifying its position, which we order by the following prescription. If $\mu < \lambda$ then $(k,\mu) < (j,\lambda)$ for all $j$ and $k$. If $\mu = \lambda$ then $(k,\mu) < (j,\lambda)$ only if $k < j$. Now we can define the spin representation of the fermion ladder operators
\begin{equation}
\label{JWtransformation}
c_{j}^{(\lambda)} = \bigotimes\limits_{(k,\mu) < (j,\lambda)}\hspace{-3mm} \sigma^{z (\mu)}_{k} \sigma^{- (\lambda)}_{j},
\end{equation}
which satisfy the required anticommutation relations by construction. 

Applying this transformation to the Hamiltonian of each chain and discarding a constant energy shift yields 
\begin{equation}
\label{spinH}
H_F^{(\lambda)} = H^{(\lambda)} + 2\Delta \sum\limits_{j=1}^{N} \sigma^{z (\lambda)}_{j} - \Delta(\sigma^{z (\lambda)}_1 + \sigma^{z (\lambda)}_{N}),
\end{equation}
where $H^{(\lambda)}$ is given by equation\ \eqref{hami}. The second term represents a homogeneous magnetic field, which makes no difference to steady-state magnetisation and current expectation values and can be thrown away \cite{we2}. The third term represents magnetic fields acting on the boundary sites, which can be neglected for large $N$. We have checked numerically that the effect of incorporating these boundary fields disappears as $N$ increases. 

Now we consider how the transformation acts on the dissipators \eqref{fermionSiteLiouvillian}. The Lindblad operators transform as 
\begin{equation}
\label{spinLindbladOps}
A_{j}^{(\lambda)} = - S_{j}^{(\lambda)} L_{j}^{(\lambda)},
\end{equation}
where the Jordan-Wigner string is defined by
\begin{equation} \label{JWstring}
 S_{j}^{(\lambda)} = \hspace{-5mm} \bigotimes\limits_{(j,\lambda) < (k,\mu) < (j,\lambda+1)}\hspace{-5mm} \sigma^{z (\mu)}_{k},
\end{equation}
while $L^{(\lambda)}_{j} = L^{(\lambda,\lambda+1)}_{j}$ is defined by equation\ \eqref{incoh_coup}. Note also that $[L_{j}^{(\lambda)} ,S_{j}^{(\lambda)}] = 0$. 

equation\ \eqref{spinLindbladOps} implies that the anticommutator terms in equation\ \eqref{fermionSiteLiouvillian} have a simple transformation, since $A^{(\lambda) \dagger}_{j}A^{(\lambda)}_{j} = L^{(\lambda) \dagger}_{j}L^{(\lambda)}_{j}$ and $A^{(\lambda)}_{j}A^{(\lambda) \dagger}_{j} = L^{(\lambda) }_{j}L^{(\lambda) \dagger}_{j}$. The ``sandwich" terms transform, for example, as
\begin{equation}
\label{sandwichTerm}
A_{j}^{(\lambda)} O A_{j}^{(\lambda) \dagger} =  L_{j}^{(\lambda)} S_{j}^{(\lambda)} O S_{j}^{(\lambda)} L_{j}^{(\lambda) \dagger}\qquad A_{j}^{(\lambda) \dagger} O A_{j}^{(\lambda)} =  L_{j}^{(\lambda) \dagger} S_{j}^{(\lambda)} O S_{j}^{(\lambda)} L_{j}^{(\lambda)}.
\end{equation}

The only observables we consider are linear combinations of the operators $n^{(\lambda)}_j = \frac{1}{2}(1+\sigma^{z (\lambda)}_j)$, $n^{(\lambda)}_jn^{(\lambda)}_{j+1}$ and $c^{(\lambda) \dagger}_{j} c^{(\lambda)}_{j+1} = -\sigma^{+ (\lambda)}_{j} \sigma^{- (\lambda)}_{j+1}$, along with their Hermitian conjugates and the identity operator. Operators that commute with $n^{(\lambda)}_{j}$ also commute with the string operators \eqref{JWstring}, which therefore disappear from equation\ \eqref{sandwichTerm} since $(S_{j}^{(\lambda)} )^2 = 1$. 

However, the hopping operators $c^{(\lambda) \dagger}_{j} c^{(\lambda)}_{j+1} $ do not commute with $n^{(\lambda)}_{j}$ and must therefore be considered in more detail. The string $S_{j}^{(\lambda)}$ contains all $\sigma^{z}$ operators acting on the sites from $(j+1,\lambda)$ to $(j-1,\lambda+1)$, inclusive. Therefore, the strings appearing in the sandwich terms of $\mathcal{A}^{(\mu) \dagger}_{k}$ commute with $c^{(\lambda) \dagger}_{j} c^{(\lambda)}_{j+1} $ unless $(k,\mu) = (j,\lambda)$ or $(k,\mu) = (j+1,\lambda-1)$. Nevertheless, in these two special cases the sandwich terms identically vanish and therefore
\begin{equation}
\label{sandwichVanish}
A_{j}^{(\lambda)}c^{(\lambda) \dagger}_{j} c^{(\lambda)}_{j+1} A_{j}^{(\lambda) \dagger} = L_{j}^{(\lambda)}c^{(\lambda) \dagger}_{j} c^{(\lambda)}_{j+1} L_{j}^{(\lambda) \dagger} = 0,
\end{equation}
and similar relations hold for the other potentially troublesome sandwich terms. 

Gathering all the results, we find that the evolution equation for each observable of interest can be written as
\begin{equation}
\label{spinHeisenbergEquation}
\frac{\partial O}{\partial t} = i \sum\limits_{\lambda = 1}^{\Lambda} [H^{(\lambda)}, O] + \sum\limits_{\lambda=1}^{\Lambda} \sum\limits_{j=1}^{N} \mathcal{L}_{j}^{(\lambda) \dagger}(O),
\end{equation}
where 
\begin{equation}
\label{spinSiteLiouvillian}
\mathcal{L}_{j}^{(\lambda) \dagger}(O) \; =\; L_{j}^{(\lambda)} O L_{j}^{(\lambda) \dagger} - \frac{1}{2} \{ L_{j}^{(\lambda)} L_{j}^{(\lambda) \dagger}, O \} + L_{j}^{(\lambda) \dagger} O L_{j}^{(\lambda)} - \frac{1}{2} \{ L_{j}^{(\lambda) \dagger} L_{j}^{(\lambda)}, O \}.
\end{equation}
Transforming back to the Schr\"{o}dinger picture, we arrive at the master equation described in Section \ref{model}. Dephasing terms can also be straightforwardly included, since the dephasing Lindblad operators $\sigma^{z (\lambda)}_{j} = 2n_{j}^{(\lambda)} - 1$ have a local representation in terms of both spins and fermions. 

\section{Analytic approach for incoherently coupled noninteracting spin chains} \label{analytic_approach}
Following the method proposed by \v{Z}nidari\v{c} \cite{znidaric2010exact,znidaric2011solvable}, we derive an analytic approximation for the state of two limit cases of the system of non-interacting ($\Delta=0$) incoherently coupled spin chains, namely for $\Lambda=2$ and $\Lambda\to\infty$. We consider first the case of two chains, and propose the following ansatz for the NESS of the entire system:
\begin{equation} \label{ansatz}
\rho=\frac{1}{2^{2N}}(I+A+B+\cdots).
\end{equation} 
Here $I$ is the identity operator of the entire system, and $A$ and $B$ are functions of spin operators in $z$ direction and spin currents, respectively:
\begin{equation}
 A=\sum_{\lambda=1}^{2}\sum_{j=1}^{N}a_{j}^{(\lambda)}\sigma_{j}^{z(\lambda)}\quad B=\frac{b}{8}\sum_{\lambda=1}^{2}\sum_{k=1}^{N-1}j_{k}^{(\lambda)}.
\end{equation} 
As seen below, $A$ and $B$ scale with $f$, so the approximation only gives the NESS up to first order in the driving strength. Nevertheless, this is enough to obtain exact results of the current and magnetisation of the chains for all values of $f$, since to obtain $r$-point correlation functions, an expansion up to order $r$ is needed \cite{znidaric2010exact}. It is easily shown that the local magnetisation and the spin current are given by
\begin{equation}
 \langle\sigma_{j}^{z(\lambda)}\rangle=\operatorname{Tr}(\rho\sigma_{j}^{z(\lambda)})=a_{j}^{(\lambda)}\qquad\langle j_{k}^{(\lambda)}\rangle=\operatorname{Tr}(\rho j_{k}^{(\lambda)})=b.
\end{equation}  
The master equation for the NESS reads
\begin{equation}
\partial\rho/\partial t=\mathcal{L}_{\text{H}}(\rho)+\mathcal{L}_{\text{driv}}(\rho)+\mathcal{L}_{\text{inc}}(\rho)=0,
\end{equation} 
where $\mathcal{L}_{\text{H}}(\rho)=-i[H,\rho]$, and $\mathcal{L}_{\text{driv}}(\rho)$ and $\mathcal{L}_{\text{inc}}(\rho)$ are the Lindblad terms corresponding to driving at the boundaries and interchain coupling, respectively, with the jump operators of equations \eqref{lind_edges_coup} and \eqref{incoh_coup}. Now we introduce the ansatz \eqref{ansatz} in the master equation. Since both chains are equal, they have the same dynamics and steady state, so we set $a_{j}^{(\lambda)}=a_{j}$. We then obtain the following results for each process:
\begin{equation}
 \mathcal{L}_{\text{H}}(\rho)=\frac{1}{2^{2N}}\sum_{\lambda=1}^{2}\Bigl(b(\sigma_{N}^{z(\lambda)}-\sigma_{1}^{z(\lambda)})+\sum_{i=1}^{N-1}(a_{i}-a_{i+1})j_{i}^{(\lambda)}\Bigr)
\end{equation} 
\begin{equation}
  \mathcal{L}_{\text{driv}}^{\text{L}}(\rho)=\frac{1}{2^{2N}}\sum_{\lambda=1}^{2}\frac{\Gamma}{2}\biggl(2f\sigma_{1}^{z(\lambda)}-2a_{1}\sigma_{1}^{z(\lambda)}-\frac{b}{4}j_{1}^{(\lambda)}\biggr)
\end{equation} 
\begin{equation}
\mathcal{L}_{\text{driv}}^{\text{R}}(\rho)=-\frac{1}{2^{2N}}\sum_{\lambda=1}^{2}\frac{\Gamma}{2}\biggl(2f\sigma_{N}^{z(\lambda)}+2a_{N}\sigma_{N}^{z(\lambda)}+\frac{b}{4}j_{N-1}^{(\lambda)}\biggr)
\end{equation}
where we divided the contribution of the driving into its left (L) and right (R) components ($ \mathcal{L}_{\text{driv}}(\rho)=\mathcal{L}_{\text{driv}}^{\text{L}}(\rho)+\mathcal{L}_{\text{driv}}^{\text{R}}(\rho)$), and
\begin{equation}
\mathcal{L}_{\text{inc}}(\rho)=-\frac{\gamma}{2^{2N}}\frac{b}{8}\sum_{\lambda=1}^{2}\sum_{k=1}^{N-1}j_{k}^{(\lambda)}.
\end{equation}
To obtain the $N+1$ coefficients $b$ and $a_{i}$ we need $N+1$ different equations, which result from equating to zero the coefficients in front of each operator. Explicitly, in front of $\sigma_{1}^{z(\lambda)}$ and $\sigma_{N}^{z(\lambda)}$ we have
\begin{align} \label{a1N}
 &\Gamma f-\Gamma a_{1}-b=0\quad\rightarrow\quad a_{1}=f-\frac{b}{\Gamma}\\
 -&\Gamma f-\Gamma a_{N}+b=0\quad\rightarrow\quad a_{N}=-f+\frac{b}{\Gamma}=-a_{1}.\notag
\end{align} 
Similarly, the coefficient in front of $j_{i}^{(\lambda)}$ is
\begin{equation} \label{ai}
 (a_{i}-a_{i+1})-\frac{b\gamma}{8}-\frac{\Gamma}{8}b(\delta_{i,1}+\delta_{i,N-1})=0.
\end{equation} 
The solution of this system of equations gives the magnetisation in the bulk ($i>1$)
\begin{equation} \label{magnetisation_d0}
 a_{i}=f-b\biggl(\frac{\Gamma}{8}+\frac{1}{\Gamma}+(i-1)\frac{\gamma}{8}\biggr),
\end{equation} 
and the spin current
\begin{equation}
 b=\frac{4f}{(\Gamma/4)+(4/\Gamma)+(N-1)(\gamma/4)}.
\end{equation} 
Note that up to $O(f)$, the total state of the system $\rho$ is a product of the states of each chain ($\rho_{1}$ and $\rho_{2}$). So, if we have
\begin{equation}
 \rho_{\lambda}=\frac{1}{2^{N}}\biggl(I+\sum_{j=1}^{N}a_{j}\sigma_{j}^{(\lambda)z}+\frac{b}{8}\sum_{k=1}^{N-1}j_{k}^{(\lambda)}\biggr),
\end{equation} 
it follows that, up to $O(f)$, $\rho=\rho_{1}\otimes\rho_{2}$ (mean-field approximation), with $\rho$ given by the ansatz of equation \eqref{ansatz}.

For the case $\Lambda\to\infty$ with homogeneous incoherent coupling $\gamma$, described in Section \ref{autocoupling} (self-coupled chain), we follow a similar process. Assuming an ansatz for the NESS of the chain like that of equation~\eqref{ansatz}, with normalisation to $2^N$, we obtain
\begin{equation}
\mathcal{L}_{\text{inc}}(\rho)=-\frac{\gamma}{2^{N}}\frac{b}{4}\sum_{k=1}^{N-1}j_{k}.
\end{equation}
The NESS is then equivalent to that of of two chains, but with $\gamma/2$ instead of $\gamma/4$ in the factors $a_i$ and $b$. \\

\begin{figure}
\begin{center}
\includegraphics{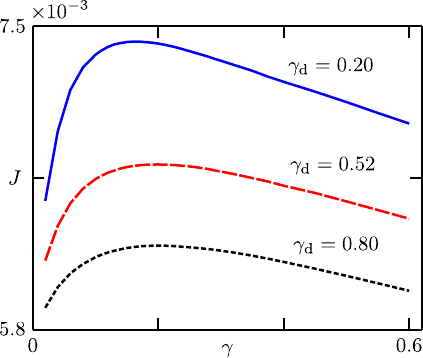}
\caption{\label{simultaneous_current} Spin current through a strongly interacting system ($\Delta=2$ and $N=40$) when both dephasing and incoherent self-coupling take place, as a function of $\gamma$, for different dephasing rates ($\gamma_{\text{d}}=0.52$ is the optimal dephasing in the absence of incoherent coupling. Note that here $\gamma_{\text{d}}$ is defined to be twice that of Ref.~\cite{we}).}
\end{center}
\end{figure}

\section{Simultaneous presence of incoherent coupling and dephasing} \label{app_simultaneous}

We have performed simulations of chains with incoherent self-coupling and dephasing at the same time, and verified that the former tends to be dominant. In figure\ \ref{simultaneous_current} we show the spin current through the system as a function of the incoherent coupling, for fixed dephasing rates. For all the cases the currents are reduced from those of $\gamma_{\text{d}}=0$, but are still larger than those of dephasing alone (compare to the inset of figure\ \ref{comparison_self_deph}). This indicates that the current enhancement induced by the incoherent coupling is still the dominant mechanism. This conclusion is reinforced by looking at the rates of energy dissipation corresponding to both the incoherent coupling (equation~\eqref{dissipation_self}) and dephasing (equation~\eqref{dissipation_dephasing}). We found that for both a large and a small dephasing rate, the amplitude of the energy dissipation rate coming from the incoherent coupling is significantly larger than that of dephasing, except for very small incoherent couplings.\\

\bibliographystyle{unsrt}

\bibliography{mybib_coupled}

\end{document}